\documentclass[preprint,12pt]{elsarticle}

\usepackage{apalike}
\let\bibhang\relax
\usepackage{amssymb}
\usepackage{booktabs}
\usepackage[margin=1in]{geometry}
\usepackage{amsmath,amsfonts,amssymb,amsthm}
\usepackage{url} 
\usepackage[numbers]{natbib}

\journal{GCCENG23}
\begin{document}

\begin{frontmatter}



\title{ESP2CS: Securing Internet of Vehicles through Blockchain-enabled Communications and Payments}


\author[inst1]{Rateb Jabbar}

\affiliation[inst1]{organization={KINDI Center for Computing Research, College of Engineering, Qatar University},
            city={Doha},
            postcode={P.O. Box 2713}, 
            country={Qatar}}

\author[inst2]{Mohamed Kharbeche}

\affiliation[inst2]{organization={Qatar Transportation and Traffic Safety Center, Qatar University},
            city={Doha},
            postcode={P.O. Box 2713}, 
            country={Qatar}}

\begin{abstract}
The burgeoning domain of the Internet of Vehicles (IoV), a subset of the Internet of Things (IoT), promises to revolutionize transportation through enhanced safety, efficiency, and environmental sustainability. By amalgamating technologies like sensors and cloud computing, the IoV paves the way for optimized traffic management, heightened vehicle safety, and the birth of novel business paradigms. However, this growth is shadowed by significant security concerns, especially in the communication and payment sectors. Addressing the pressing need for secure Vehicle-to-Everything (V2X) communications and payments amidst rising cyber threats, this research introduces the Ethereum-based Secure Payment and Communication Solution (ESP2CS). Utilizing Ethereum as a middleware, ESP2CS ensures robust and secure V2X interactions. The solution is complemented by an Android Auto application for vehicles, streamlining inter-vehicle communication, parking space detection, and transaction management. Furthermore, dedicated Android applications are developed for parking space renters and the parking IoT system. Preliminary evaluations underscore ESP2CS's superior cost-effectiveness, integrity and consistency over contemporary solutions, with Ethereum bolstering both security and efficiency.
\end{abstract}


\end{frontmatter}


\section{Introduction} In the face of the exponential surge in vehicular numbers, contemporary transportation solutions, despite their advancements, grapple with the burgeoning challenges of traffic management. The imperative for the integration of Intelligent Transportation Systems (ITS) has never been more pronounced. ITS, with its potential to augment traffic efficiency and foster the development of smart roads, stands as a beacon of innovation in this context \cite{Automotivestatistics}. 
The transition from Vehicular Ad-Hoc Networks (VANETs) to the Internet of Vehicles (IoV) represents a substantial advancement towards realizing the goals of Intelligent Transportation Systems (ITS). This progression enables enhanced information exchange, efficiency, and paramount safety between vehicles and infrastructural entities, underscoring the pivotal role of interconnected vehicular communication in modern transportation systems \cite{overviewofinternetofvehicles}. According to the US Department of Transport (DOT) \cite{lamssaggad2021survey}, the IoV holds the promise of substantially reducing crashes involving unimpaired drivers. With the seamless integration of IoV, an estimated 79\% of such crashes are projected to be avoidable, courtesy of the enhanced communication and collaboration among vehicles.

Industry forecasts predict continued growth in connected vehicles, creating a profitable market.
The size of the global connected car fleet in 2021 is estimated to be 237 million units, a number projected to soar to 400 million units by 2025, 600 million units by 2030, and 1.1 billion units by 2035 \cite{SizeconnectedCar}. This burgeoning growth underscores the escalating reliance on and the integration of technology within the automotive sector. The worldwide market for connected cars is anticipated to witness a robust growth, expanding at a CAGR of 22.7\% from 2022 to 2031. According to a study by Transparency Market Research \cite{connectedcarmarket}, the market is projected to reach a substantial valuation of US\$228.3 billion by 2031. For the year 2023, the connected car market is estimated to attain a value of US\$44.4 billion.

However, the journey towards the realization of a fully integrated and secure IoV is fraught with challenges, primarily centred around ensuring the security and privacy of the diverse participants in intelligent transportation systems. The burgeoning connectivity, while offering numerous advantages, also opens the door to potential malicious attacks, making it imperative to ensure the robustness of the security infrastructure in place. This vulnerability was glaringly exposed in several incidents that underscore the criticality of robust security measures. 
A notable security breach transpired in Australia \cite{troyhunt}, where Troy Hunt, a distinguished Microsoft Security Professional, successfully executed an attack. This breach allowed him to acquire remote control over the operations of Nissan LEAFs. This incident glaringly exposed the vulnerability of interconnected vehicles to unauthorized and potentially harmful remote interventions. Further emphasizing the vulnerability, Keen Security Lab \cite{tesla}, a renowned Chinese cybersecurity firm, managed to remotely manipulate specific components of a Tesla Model S, including its brakes and mirrors, from a considerable distance of around 20 kilometres.

These cumulative incidents underscore the urgent and imperative need to augment the security of data transmission, calling for the innovation and implementation of more sophisticated and robust solutions. This includes the integration of secure payment systems to bolster the overall efficiency, security, and integrity of Internet of Vehicles (IoV) systems. As the IoV landscape continuously evolves, it is paramount to steadfastly enhance the security infrastructure, ensuring the safeguarding of all stakeholders and the uninterrupted and secure functioning of intelligent transportation systems.

In the current technological milieu, emergent technologies such as Artificial Intelligence (AI) \cite{jabbar2020driver,abdelhedi2020prediction,jabbar2022recent,abdelhedi2023machine,abulibdeh2022impact,abdelhedi2020ultrasonic,ayedi2023ai} and Blockchain are garnering significant attention for their potential to address real-world challenges. Blockchain \cite{zheng2018blockchain,rateb2021blockchain,krichen2022blockchain}, in particular, stands out as a symbol of robust security and unwavering reliability. It finds extensive application across a myriad of sectors such as healthcare \cite{jabbar2020blockchainhealthcare}, cybersecurity \cite{moulahi2023privacy}, and transportation \cite{jabbar2022noora,jabbar2020adopting,abulibdeh2022impact,jabbar2020formal,jabbar2020model}, significantly bolstering the security framework of Internet of Things (IoT) applications by facilitating the secure exchange of data sets.
The amalgamation of Blockchain technology with the Internet of Vehicles (IoV) heralds a new era of enhanced security and advanced functionalities. This integration is poised to substantially augment the security infrastructure, bolster intelligence, expand big data storage capacities, and streamline the management processes within the IoV \cite{rateb2021blockchain2}.

In light of these considerations, this study introduces the Securing Internet of Vehicles through Blockchain-enabled Communications and Payments (ESP2CS), a pioneering solution leveraging Ethereum to ensure secure Vehicle-to-Everything (V2X) communications and payments within the ITS. This innovative approach aims to augment the utilization of urban parking spaces, alleviate traffic congestion, and minimize fuel consumption and time, thereby contributing to the overarching goals of sustainability and efficiency in urban transportation.

The remainder of this paper is structured as follows: Section II provides a comprehensive review of existing literature, focusing on communication and payment systems for parking spaces that leverage Internet of Things (IoT) and Blockchain technologies. Section III delineates the architecture and characteristics of the proposed system. Section IV offers a concise summary of the testing methodology and the corresponding results. Finally, Section V concludes the paper, offering insights and potential avenues for future research in this domain.

\section{Literature Review}
In the context of securing payment and communication between different participants
of intelligent transportation systems using Blockchain, several pertinent studies have been conducted, each contributing valuable insights and highlighting various challenges and opportunities.

Ramaguru et al. \cite{ramaguru2019blockchain} devised a method employing real-time Blockchain technology to validate credentials and ensure secure vehicle connectivity. The approach leverages a distributed identifier to attain pseudo-anonymity, enhancing the security of communications. However, a significant limitation is noted: the system's functionality is compromised if authentication fails, posing a potential risk to the secure communication between vehicles.

Jiang et al. \cite{jiang2018blockchain} presented an architecture for transferring data from the automobile Blockchain to the external world, offering insights into IoV servers and their potential applications in Blockchain. This research opens new avenues for exploring IoV Blockchain technology and the possibility of implementing a multi-Blockchain architecture. Despite these advancements, the system does not support key-used vehicles, limiting its applicability in broader contexts.

Maglaras et al. \cite{maglaras2016social} explored the concept of the Social Internet of Vehicles (SIoV), a platform facilitating the connection between drivers and vehicles. The study provides a comprehensive analysis of SIoV, including its techniques, components, and potential challenges related to user privacy protection and trust in applications. The SIoV platform is recognized for its high performance, open architecture, and dynamic vehicle architecture. However, concerns regarding the framework's consistency and reliability are raised, indicating areas for further enhancement and development.

Jabbar et al. \cite{jabbar2020blockchain} developed a decentralized framework based on Blockchain technology (DISV), the DISV \cite{jabbar2020cadre} represents a real-time application specification that provides secure communication among all participants in the transportation system. The developed solution is composed of three layers: perception, network, and application layers. The perception layer assumes the form of an Android application including two sub-systems. The vehicle data collection system \cite{jabbar2018applied}, \cite{jabbar2019urban} is a sub-system for obtaining information regarding the journey and vehicle . The second sub-system, called driver drowsiness detection \cite{jabbar2018real}, \cite{jabbar2020driver}, aims to sense driver drowsiness by acquiring driver behavior information. 
This solution lacks scalability as it can not be used by a large
number of vehicles in the same region.
In addressing this issue, authors \cite{jabbar2021blockchain} introduce a novel Blockchain framework for vehicle communication and parking payment, termed PSEV . The presented framework utilizes Ethereum to forge a solution designed to enhance Vehicle-to-Everything (V2X) communications and parking payments. Additionally, the proposed solution incorporates Android auto and application modules to automate the communication process efficiently. Experimental evaluations were conducted to examine its computational costs, communication expenditures, and real-time attributes (RTA).

Zhang et al. \cite{zhang2019blockchain} examined the challenges associated with identifying a trustworthy entity for message storage and transfer in vehicle communication systems. The study emphasizes the importance of privacy maintenance while implementing countermeasures, contributing to the enhancement of secure communication. However, the proposed system demands more execution time, which may impact the real-time communication requirements of vehicle systems.

In summary, while various studies have made significant strides in enhancing the security of payment and communication between vehicles using Blockchain technology, challenges such as authentication failures, system limitations, reliability concerns, and execution time requirements persist. These insights underscore the need for continued research and innovation to develop robust, efficient, and secure Blockchain-based solutions for vehicle communication and payment systems.

\section{Proposed Solution}

\begin{figure*}[h]
	\centering
	\includegraphics[width=\linewidth]{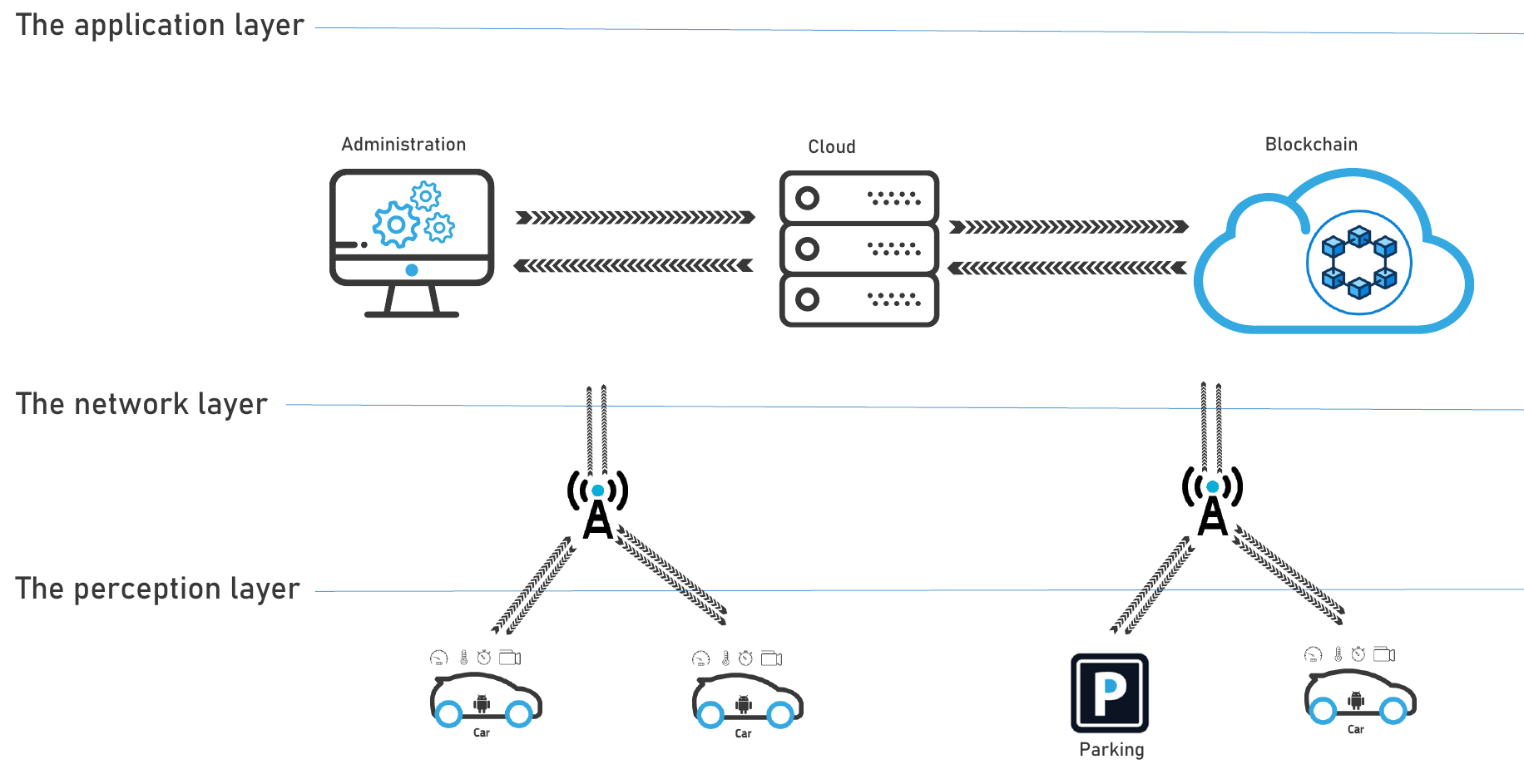}
	\caption{The architecture of the Ethereum-based Secure Payment and Communication Solution (ESP2CS).}
	\label{fig:Architecture}
\end{figure*}

To address the challenges of securing communications and payments in IoV systems, this research proposes the Ethereum-based Secure Payment and Communication Solution (ESP2CS). The decentralized architecture of ESP2CS eliminates central points of failure and establishes trust among entities through Blockchain's transparency and immutability properties.

\section{System Architecture}

The ESP2CS architecture is composed of three distinct layers as illustrated in Fig.~\ref{fig:Architecture}: perception, network, and application, mirroring the established IoV models. Each layer plays a crucial role in ensuring the robustness and efficiency of the system.
This research presents an Internet of Vehicles (IoV) framework with the objective of fostering secure communications and financial transactions within transportation systems. A comprehensive elucidation of the proposed architecture will be provided in the subsequent section.

\subsection{Perception Layer}

This layer is inclusive of three individualized applications explicitly designed for vehicles and parking infrastructure, essential for assuring secure communication and seamless financial transactions. It integrates the Vehicle Interaction Application, the Parking Space Allocator Application, and the IoT Parking Manager Application. These comprehensive applications facilitate users to a) commence communication, b) manage their allocated parking space, and c) efficiently process payments.

\subsubsection{Vehicle Interaction Application}
\begin{figure}[t!]
	\centering
	\includegraphics[width=\linewidth]{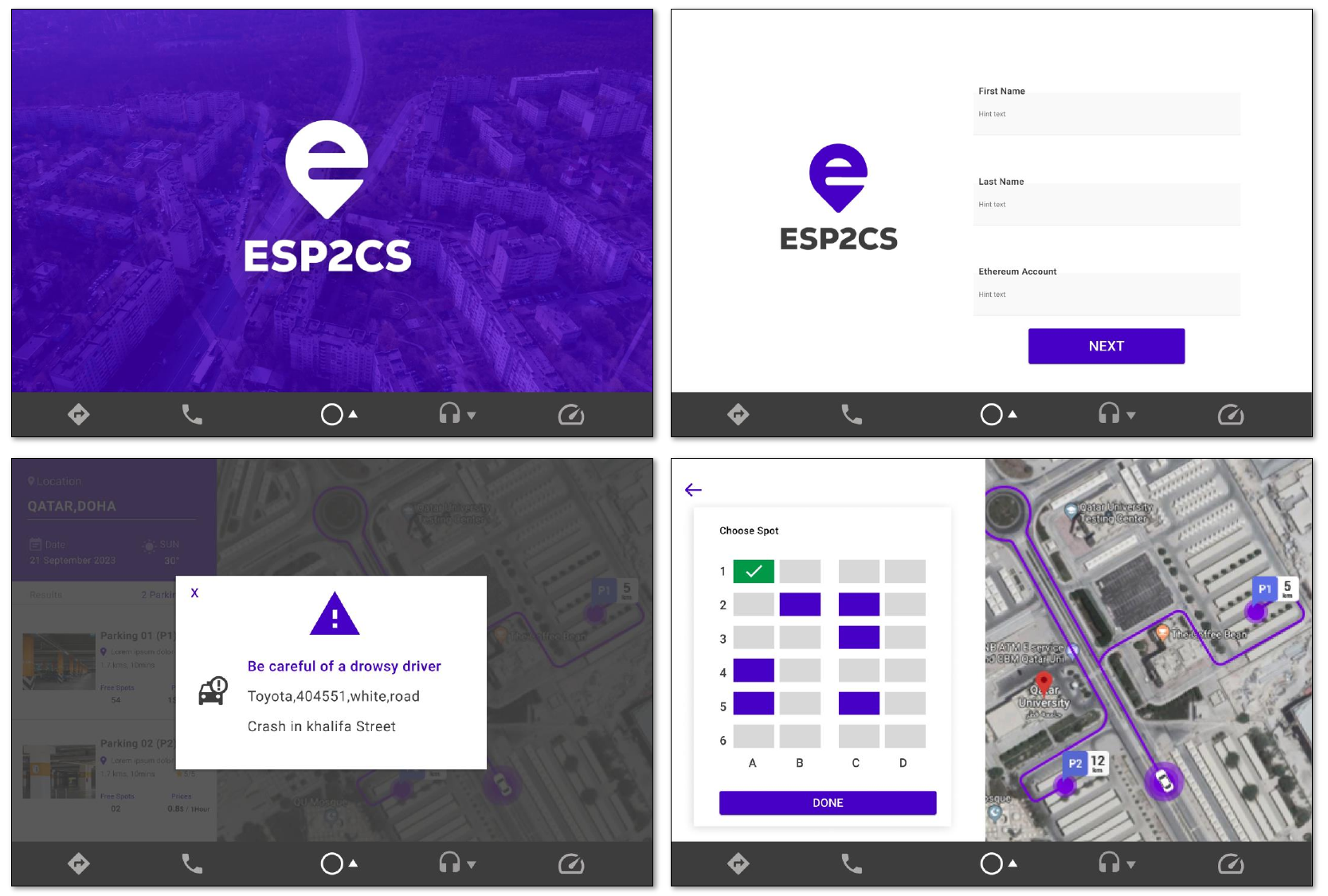}
	\caption{Screenshot of the Android application interface of the vehicle Interaction Application.}
	\label{fig:LandingPage}
\end{figure}

This tailored application serves as the vehicle interface, utilized by users for unimpeded transmission and reception of messages, alongside adept searching, navigating, and transacting for parking services. Additionally, this sophisticated application meticulously detects the vehicle's proximity to the designated parking payment station, initiating the Blockchain-based transaction therein. The application interface, as depicted in Fig.~\ref{fig:LandingPage}, incorporates an initial login page, followed by a detailed map showcasing nearby parking locales. Upon the selection of a suitable location, the application proffers comprehensive navigation information and, upon destination arrival, enumerates available parking slots along with their specific time allocations. Post the selection process, the application activates a countdown timer, which the user can manually cease upon departing from the parking, subsequently presenting a succinct transaction overview and the exact parking fee. Users are also accorded the facility to access and review their historical transactions and related details within the application.
\subsubsection{Parking Space Allocator Application}
This innovative application is a valuable asset for proprietors of buildings or parking spaces, enabling them to proficiently administer and designate time slots for their available parking spaces for public utilization. This strategic allocation allows proprietors to lucratively monetize unoccupied parking spaces. The application predominantly features a user authentication login page, a specialized parking slot configuration interface (enabling the effective allocation of time slots for parking space availability), and a comprehensive settings interface for holistic account and administrative task management. Renters can conveniently access a detailed list of past transactions alongside the total accumulated revenue for the given parking space within the application interface.
\subsubsection{IoT Parking Manager Application}
This application, embedded within the IoT parking payment apparatus, proficiently identifies proximate vehicles and authorizes access to parking slots upon receipt of driver requests. This advanced payment portal seamlessly manages authentication protocols, efficient time slot allocation, and smooth payment transactions between vehicles and parking space providers, assuring unhindered physical access to the parking facilities.
\subsection{Network Layer}

The network layer is pivotal for establishing and managing connections between diverse devices, servers, and vehicles. It employs advanced communication technologies, including 3G/4G/5G and WiFi, to ensure uninterrupted and high-speed connectivity.

\subsection{Application Layer}

At the core of ESP2CS lies the application layer, housing the Ethereum Blockchain backend and a central cloud server. The Blockchain component is instrumental for secure V2X communications and parking payments, enabled through meticulously crafted smart contracts. This layer guarantees a tamper-proof record of all transactions and messages, bolstering the system's security and reliability.

\subsubsection{Core Cloud Server}
The main cloud server functions as the host for managerial and administrative software and applications that are essential for the operation of the service. Its role involves continuous supervision of the systems to ensure the intact functionality of the entire infrastructure. Hosting the service, it stands as the chief administrative connection for all the payment gateways and vehicles within the Blockchain. This server proactively sends out communication invites to IoT devices in proximity, facilitating their interaction through existing Blockchain cloud instances.

A specific web application, as depicted in Fig.~\ref{fig:Architecture}, is stationed within the central cloud. This application acts as an instrumental tool in the application layer, dedicated to fostering interactions among diverse IoT solution components. These include the administrative web platform, a device database server that liaisons with the Blockchain, and the integral systems at the payment portals. The service is hosted using the Windows Azure cloud service paired with REST API calls. This web application is also equipped with multiple scientific tools designed to examine traffic flow at various parking locations, thereby aiding the understanding and analysis of parking demand in diverse city locales. Hence, it acts as a unified administrative platform, enabling administrators to efficiently address system issues and ensuring uninterrupted operation.

\subsubsection{Blockchain layer}
The Blockchain framework holds the responsibility of enabling communication between different elements and roadway users within ITS and managing the payment transactions between vehicles and parking services. Specifically, any intended parking payment transaction is broadcasted to the Blockchain layer for validation. Once verified, the transactions are permanently recorded as a part of a block, affirming their validity and unchangeability.

In unison with the Android application, the Blockchain layer contributes to the development of decentralized applications (DApp). These decentralized applications operate over a distributed Internet platform functioning on a peer-to-peer decentralized network such as Blockchain. In this setup, the Android application forms the front-end, while the Blockchain framework constitutes the back-end portions of the decentralized applications. Smart contracts deployed on each Ethereum node play a crucial role, with each contract serving unique functions to ensure fluid and secure operations. The mobile application interacts with these contracts, utilizing the Blockchain for sending messages, all communicated between the mobile and node-endpoint. This advanced solution incorporates the robust Web3.Js framework for Android, ensuring reliable operations.

In essence, every smart contract within the system holds significant importance, each dedicated to ensuring the smooth, secure, and efficient operation of the entire ecosystem. Here is a description list of some of the most crucial smart contracts within the system:

\paragraph{Vehicular Communication} \
\\

The “Vehicular Communication'' contract is instrumental in augmenting vehicular communication within the ESP2CS system. It embodies two main functions. The “publishMessage'' function is adeptly designed to streamline the process for vehicles to broadcast messages, enhancing the robustness of inter-vehicular communication. It meticulously stores the message content, sender's address, and timestamp, thereby fortifying the integrity and traceability of messages. Concurrently, the “readMessage'' function stands as a cornerstone for ensuring the secure and transparent retrieval of messages, permitting entities to access messages using unique IDs, and safeguarding their confidentiality and integrity. The “sendMessage'' function enhances secure and efficient vehicular communication by permitting the dispatch of messages to designated recipients, augmenting the reliability and timeliness of communication. The “getUnreadMessages'' and “markAllAsRead'' functions contribute significantly to efficient message management, allowing users to seamlessly retrieve unread messages and mark all messages as read.

\paragraph{Payment Management}\
\\
The “Payment Management'' contract stands as a linchpin for overseeing payments within the ESP2CS ecosystem, bolstering the security and transparency of financial transactions. The “makePayment'' function is intricately crafted to facilitate users in executing payments to the contract, ensuring the legitimacy and financial security of transactions. The “requestRefund'' function empowers users to seamlessly request refunds, maintaining the integrity of financial transactions by validating the refund amount. The “processRefund and “withdrawFunds'' functions further augment the financial operations, enabling the contract owner to process refunds and withdraw funds transparently and efficiently.
\paragraph{Parking Space Management}\
\\
The “Parking Space Management'' contract is a pivotal element in enhancing the efficiency and transparency of parking space management within the ESP2CS system. The “registerParkingSpace'' function enables the effortless registration of new parking spaces, ensuring optimal parking space allocation and availability. The “bookParkingSpace'' and “isAvailable'' functions synergistically ensure the efficient booking of available parking spaces, optimizing space utilization and enhancing user experience. The “releaseParkingSpace'' function guarantees the timely release of booked parking spaces, fostering effective space management and availability.

\paragraph{Automated Parking Payments}\
\\
The “Automated Parking Payments'' contract is vital for the automation of parking payments within the ESP2CS system. The “registerParkingSpace'' and startParking'' functions facilitate the effortless registration of parking spaces and initiation of parking sessions, enhancing the efficiency and transparency of the parking payment system. The “endParking'' function adeptly manages the termination of parking sessions and handles payments, ensuring transactional security and transparency. The “calculateParkingFee'' and “checkAmountDue'' functions offer an efficient mechanism for fee calculation and amount due checking, further enhancing financial management and transparency within the system.

\subsection{Optimized Consensus Protocol}
To achieve low latency in V2X communications, ESP2CS incorporates a lightweight proof-of-authority (PoA) consensus that restricts mining privileges to a set of trusted nodes \cite{zohar2015bitcoin}. This ensures faster block confirmations compared to Ethereum's computationally intensive proof-of-work. The cloud servers act as the authorities running the PoA protocol.
\subsection{Lightweight Vehicle Application}
To improve efficiency for resource-constrained vehicles, the Android vehicle application in ESP2CS employs a light client that syncs periodically with the Blockchain instead of participating in demanding mining activities \cite{leiding2016self}. The client submits transactions to cloud servers, which interact with the Blockchain on its behalf.

\section{System Analysis and Performance}
The performance and security analysis of ESP2CS is presented based on the following criteria:

\subsection{Cost Analysis}

\begin{figure*}[h]
	\centering
	\includegraphics[width=\linewidth]{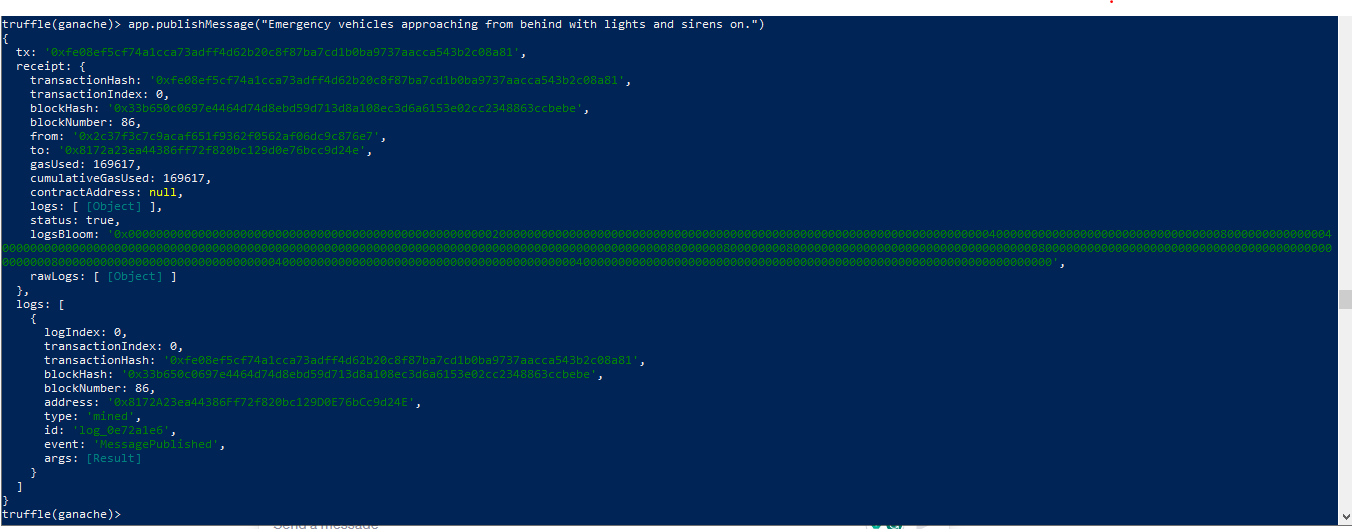}
	\caption{Screenshot of the result from invoking the ``publishMessage'' function of the ``VehicularCommunication'' smart contract}
	\label{fig:ExamplePublishMessage}
\end{figure*}

In the comprehensive evaluation of the proposed ESP2CS system, a detailed and systematic approach is employed to accurately measure the gas consumption associated with the execution of each smart contract function. The smart contracts, pivotal to the functionality of the ESP2CS system, are deployed on a local Ethereum Blockchain node. This controlled environment facilitates the precise analysis of gas usage for each function call, simulating diverse operational scenarios that mirror real-world applications. The Ethereum framework inherently documents the gas expended for each transaction, ensuring a detailed and exact capture of gas consumption metrics as illustrated in Fig.~\ref{fig:ExamplePublishMessage}. This data undergoes a thorough analytical review, providing valuable insights into the gas efficiency of each component within the ESP2CS system and highlighting potential areas for optimization, thereby enhancing the system's overall efficiency and cost-effectiveness as shown in Table \ref{tab:smart-contract-functions-rounded}.

In the Ethereum Blockchain, the computational effort required for executing transactions and interacting with smart contracts is quantified as ``gas.'' This gas is denominated in Ether (ETH), the native cryptocurrency of the Ethereum network \cite{etherscan}. The total expenditure in ETH for a transaction or smart contract operation is computed by multiplying the amount of gas consumed by the gas price, which is established at the time of transaction initiation. Mathematically, the total cost (\(C\)) in ETH is expressed as:
\begin{equation}
    C = G \times P \times 10^{-9}
\end{equation}
where \(C\) is the total cost in ETH, \(G\) is the gas used for the transaction, and \(P\) is the gas price in Gwei (Gigawei), where \(1 \, \text{Gwei} = 1 \times 10^{-9} \, \text{ETH}\). This formula calculates the total cost in ETH by taking the product of the gas used and the gas price (in Gwei), and then multiplying the result by \(10^{-9}\) to convert the cost to ETH. The gas price is typically set by the user initiating the transaction, and it plays a crucial role in determining the priority of the transaction in the network. Transactions with higher gas prices are generally processed more quickly by the network, ensuring timely and efficient operations within the ESP2CS system.

\begin{table}[h]
    \centering
    \begin{tabular}{@{}llrr@{}}
        \toprule
        \textbf{Smart Contract} & \textbf{Function} & \textbf{Gas} & \textbf{Cost (USD)} \\
        \midrule
        PaymentManagement & makePayment & 43608 & 0.555 \\
                            & requestRefund & 93433 & 1.190 \\
                            & processRefund & 0 & 0 \\
                            & withdrawFunds & 34812 & 0.443 \\
        \addlinespace
        ParkingSpaceManagement & registerParkingSpace & 88361 & 1.125 \\
                                & bookParkingSpace & 93903 & 1.196 \\
                                & isAvailable & 0 & 0 \\
                                & releaseParkingSpace & 14675 & 0.187 \\
                                & withdraw & 28771 & 0.366 \\
        \addlinespace
        VehicularCommunication & SendMessage & 168989 & 2.151 \\
                                & getUnreadMessages & 0 & 0 \\
                                & markAllAsRead & 22986 & 0.293 \\
                                & publishMessage & 169617 & 2.159 \\
                                & readMessage & 0 & 0 \\
        \addlinespace
        AutomatedParkingPayments & registerParkingSpace & 105078 & 1.338 \\
                                  & startParking & 75940 & 0.967 \\
                                  & endParking & 42594 & 0.542 \\
                                  & calculateParkingFee & 25468 & 0.324 \\
                                  & checkAmountDue & 16835 & 0.214 \\
                            
        \bottomrule
    \end{tabular}
    \caption{ Cost for ESP2CS Smart Contract Functions}
    \label{tab:smart-contract-functions-rounded}
\end{table}

\subsection{Integrity}
Robust data integrity mechanisms are essential for secure sharing of sensitive information in the proposed decentralized system. Integrity ensures data remains demonstrably accurate, consistent, and trustworthy throughout its entire lifecycle \cite{wei2020blockchain}. The system leverages Blockchain's cryptographic proofs, Merkle tree data structures, and Byzantine fault-tolerant consensus to provide strong technical protections for integrity.

Integrity best practices dictate compliance with principles of:

\begin{itemize}
	\item[-] Precision - Data is validated through multi-node computational consensus to detect errors.
	\item[-] Non-repudiation - Digitally signed chain of custody preserves originality and sources.
	\item[-] Chronometric accuracy - Distributed ledger timestamps data in real-time at observation.
	\item[-] Accessibility - Data is permanently recorded in standardized formats.
    \item[-] Auditability - Append-only ledger logs all changes with setter metadata.
\end{itemize}

Merkle trees \cite{merkle1980protocols}, a fundamental component in Blockchain technology, use hash pointers for efficient keyed-hash message authentication of data chunks, enabling lightweight proofs of validity and minimizing storage overhead. Their succinct cryptographic commitments facilitate transmitting compact integrity verification across peer networks.
Hash-chained Blockchain structures match data blocks to unforgeable cryptographic fingerprints, ensuring tamper-evidence despite encryption \cite{nakamoto2008bitcoin}. Digital signatures further validate modifications, with version histories proving provenance.
Ongoing zero-knowledge proof research expands integrity assurances through validity demonstrations without revealing data contents \cite{goldwasser1989knowledge}. Statistical audits will quantify reliability rates over time using randomness tests. Formal analysis of Merkle tree game theory security will prove resilience to attack vectors.
\\In essence, the ESP2CS system, leveraging advanced Blockchain technology that use defense-in-depth approach synergistically combines decentralized consensus, cryptographic commitments, and immutable data structures to assure high integrity of sensitive shared data
\subsection{Confidentiality}
In the intricate landscape of computer systems, the essence of confidentiality is paramount, resonating with the assurance that access to sensitive and protected data is meticulously restricted to authorized entities. The ESP2CS system, rooted in Blockchain technology, embarks on a comprehensive approach to fortify the confidentiality of data and transactions, safeguarding them against unauthorized access and potential malicious endeavors.
\\In the Blockchain domain, confidentiality transcends the conventional boundaries, ensuring not only the secure execution of transactions but also the protection of transaction details and participant identities. Despite the inherent transparency of public Blockchains, the ESP2CS system adeptly navigates this terrain to bolster confidentiality. It employs mechanisms akin to those in private Blockchains, ensuring the concealment of transaction information and participant identities, thereby enhancing the security and privacy dimensions of the system.
\\The system is meticulously designed to meet stringent confidentiality requirements, ensuring that non-participating entities are unable to access transaction details unless explicitly shared by the participating parties. This robust framework further extends to the protection of participant identities, preventing unauthorized third-party access and ensuring the seamless and secure execution of transactions within the ESP2CS ecosystem. The strategic implementation of these confidentiality-enhancing mechanisms underscores the ESP2CS system's commitment to providing a secure, private, and robust platform for the execution of transactions, reinforcing its position as a reliable and trustworthy solution in the Blockchain space.
\subsection{Consistency}
The Ethereum Blockchain consensus mechanism establishes trust in evaluation results without the need for manual reconciliation of divergent assessments \cite{lof2017decentralized}. This consistency mechanism relies on honest nodes always building on the longest chain, the one containing the most consensus protocol. Even when nodes temporarily disagree on performance ratings due to network latency, the decentralized peer-to-peer structure ensures they rapidly converge on consistent truths as communications resume.
\\By cryptographically accepting blocks, nodes intrinsically reject alternative forked histories that could undermine integrity with contradictory performance results. The Blockchain's inherent resistance to manipulation thus bolsters confidence in the consistency of ratings. Ongoing scrutiny of consensus algorithms will be essential to ensure continued integrity as the system scales. Quantitative benchmarking of agreement rates over time and across random test samples will confirm the Blockchain enables consistent evaluation.
\\The Blockchain's tamper-proof ledger of performance ratings provides a trustworthy foundation for analysis. Nevertheless, corrupted nodes could still compromise integrity. Best practices including cryptography, redundancy, and Byzantine fault tolerance techniques \cite{castro1999practical} will harden system nodes against bad actors. Continued research into advanced consensus protocols and defense-in-depth strategies will further bolster the consistency and integrity of performance evaluation. Taken together, the decentralized, transparent Blockchain architecture intrinsically promotes trustworthy and consistent system performance assessment essential for robust validation.

\section{Conclusion}
This research presented ESP2CS, an Ethereum Blockchain-based framework tailored to address security and privacy concerns surrounding communications and payments in Internet of Vehicles systems. The decentralized architecture of ESP2CS provides enhanced resilience compared to centralized architectures by eliminating single points of failure.

Encrypted communications over Blockchain ensure privacy-preserving messaging between vehicles and infrastructure. For parking payments, ESP2CS enables automated transactions via smart contracts triggered based on vehicle proximity. A lightweight vehicle application design reduces computational overhead for resource-constrained vehicles.

Experimental evaluations demonstrate the cost-effectiveness, integrity and consistency of ESP2CS in IoV environments. The immutability and transparency properties of the Ethereum Blockchain establish trust and accountability for V2X interactions. ESP2CS thus provides a promising approach for developing secure and efficient next-generation intelligent transportation systems and autonomous vehicles reliant on seamless connectivity.
As future work, additional enhancements can be incorporated into ESP2CS, like integration with e-governance frameworks for vehicular regulations and support for micropayment channels to further improve efficiency and scalability.

\newlength{\bibhang}
\setlength{\bibhang}{12pt} 
\bibliography{references.bib}
\bibliographystyle{IEEEtranN}
\end{document}